\pdfoutput=1
\documentclass[preprint2]{aastex631}
\usepackage{natbib}
\usepackage{hyperref}
\usepackage{float}
\usepackage{amsmath}
\usepackage[utf8]{inputenc}
\usepackage{graphicx}
\usepackage{listings}
\usepackage[caption=false]{subfig}
\usepackage{nccfoots}

\begin{document}

\title{Searching for Transit Timing Variations and Fitting a New Ephemeris to Transits of TrES-1 b} 
\date{June 2022}

\author[0000-0001-9968-1069]{Paige Yeung}
\author[0000-0003-3085-2178]{Quinn Perian}
\affil{Massachusetts Institute of Technology, 77 Massachusetts Ave, Cambridge, MA 02139}
\author{Peyton Robertson}
\affil{Stanford University, 450 Serra Mall, Stanford, CA 94305}
\author[0000-0001-6554-1826]{Michael Fitzgerald}
\affil{Las Cumbres Observatory, 6740 Cortona Dr Suite 102, Goleta, CA 93117-5575, USA}
\author{Martin Fowler}
\affil{Citizen Scientist, Les Rocquettes, Orchard Road, South Wonston, Winchester SO21 3EX, UK}
\author{Frank Sienkiewicz}
\affil{Harvard-Smithsonian Center for Astrophysics, Smithsonian Astrophysical Observatory, 60 Garden Street, MS-71, Cambridge, Massachusetts 02138}
\author{Kalée Tock}
\affil{Stanford Online High School, 415 Broadway Academy Hall, Floor 2, 8853, Redwood City, CA 94063}


\begin{abstract}
    Based on the light an exoplanet blocks from its host star as it passes in front of it during a transit, the mid-transit time can be determined. Periodic variations in mid-transit times can indicate another planet's gravitational influence. We investigate 83 transits of TrES-1 b as observed from 6-inch telescopes in the MicroObservatory robotic telescope network. The EXOTIC data reduction pipeline is used to process these transits, fit transit models to light curves, and calculate transit midpoints. This paper details the methodology for analyzing transit timing variations (TTVs) and using transit measurements to maintain ephemerides. The application of Lomb-Scargle period analysis for studying the plausibility of TTVs is explained. The analysis of the resultant TTVs from 46 transits from MicroObservatory and 47 transits from archival data in the Exoplanet Transit Database indicated the possible existence of other planets affecting the orbit of TrES-1 and improved the precision of the ephemeris by one order of magnitude. We now estimate the ephemeris to be
 $(2455489.66026 \text{ BJD}_\text{TDB} \pm 0.00044 \text{ d}) + (3.0300689 \pm 0.0000007)\text{ d} \times \text{epoch}$. This analysis also demonstrates the role of small telescopes in making precise mid-transit time measurements, which can be used to help maintain ephemerides and perform TTV analysis. The maintenance of ephemerides allows for an increased ability to optimize telescope time on large ground-based telescopes and space telescope missions.
\end{abstract}
\keywords{ephemerides, methods: data analysis, planets and satellites: detection, techniques: image processing}
\section{Introduction}
The transit method is used to determine the radius of a planet and its orbital period by measuring the \textcolor{black}{decrease in stellar flux} when the planet blocks its host star from the \textcolor{black}{line of sight of the observer}. Many transiting exoplanets observed by NASA's Kepler and \textcolor{black}{Transiting Exoplanet Survey Satellite (TESS)} missions are ideal for atmospheric characterization \citep{ETS}. However, transit midpoint uncertainties increase over time. A network of small telescopes used by citizen scientists can be used to maintain ephemerides by perform such regular observations fairly accurately \citep{Fowler}. As a result, small telescopes can help save time on larger telescopes by continually observing transits and keeping their parameters up to date \citep{ETS}.

A meta-analysis of a collection of transit measurements can also provide useful information. Such collections are useful for studying transit timing variations (TTVs), represented by a difference between the expected (or calculated from a least-square regression) and observed transit midpoints of a planet (\citealt{corteszuleta2020tramos}, \citealt{Agol2018}). These variations might constitute the signature of another planet in the system. Collecting and assimilating data from multiple transits can be a way to identify otherwise undetected planets in the system.

In this paper, transits of TrES-1 b, an exoplanet that is part of the class of gas giants known as ``hot Jupiters,'' are analyzed to explore the potential of TTVs and update the ephemeris. TrES-1 b was first discovered in 2004 by the Trans-Atlantic Exoplanet Survey (TrES) using the transit method. The star that TrES-1 b orbits was measured to have a V-band magnitude of $11.790$ \citep{Alonso_2004}.

\section{Data Collection and Reduction}

Observations of 83 transits of TrES-1 b, collected between 2010 and 2020, were analyzed. These transits were observed by Cecilia, a 6-inch telescope from the MicroObservatory Robotic Telescope Network \citep{MicroObservatoryNet}\footnote{\textcolor{black}{https://mo-www.cfa.harvard.edu/MicroObservatory/}}. All of the images of TrES-1 b where taken from the Fred Lawrence Whipple Observatory in Arizona, where Cecilia is currently installed. Each image had a field of view that is $60$ arcminutes horizontally and $40$ arcminutes vertically. Each exposure time was 60 seconds. The focal length of the telescope was 560 mm, and no filter was used. The telescope's main camera uses a Kodak KAF-1402ME image sensor with resolution $1000\times 1400$ and $6.8\mu \text{m}$ square pixels with a resolution of 2.5 arcminutes per pixel. The sensor is cooled by a two-stage
Peltier thermo-electric cooler \citep{MicroObservatoryNet}.

To process these data, the EXOplanet Transit Interpretation Code (EXOTIC) pipeline\footnote{\textcolor{black}{https://github.com/rzellem/exotic}} was used to find the mid-transit time and uncertainty for each set of time series observations \citep{ETS}. In the EXOTIC pipeline, the images were first reduced by performing dark subtraction using the darks we provided. The pixel coordinates of the target star and comparison stars (comp stars), if any were used, were then detected.  EXOTIC tracked the target star across the images based on these initial pixel coordinates.  For each image, it computed the total flux of the star within an aperture and subtracted the average flux over a sky background annulus for an equivalent number of pixels \citep{ETS}.  It selected the aperture radius, sky background annulus, and the best of the detected comp stars by minimizing the residual root-mean-square (RMS) scatter of a \cite{MandelAgol} model transit curve. It fit the resulting light curve parameters using a Markov Chain Monte Carlo (MCMC) fitting routine \citep{Ford_2006}.

\begin{figure}
    \centering
    \includegraphics[width=0.45\textwidth]{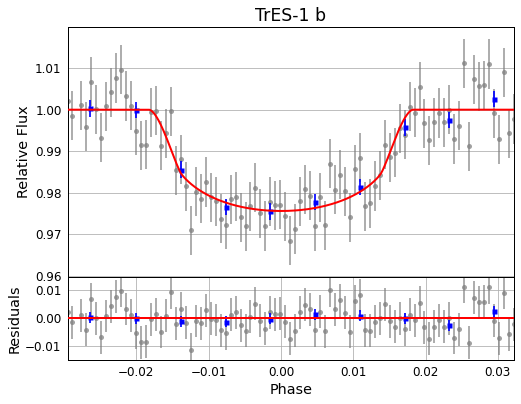}
    \caption{Light curve fit by EXOTIC for the transit of TrES-1 b on May 25, 2011.}
    \label{fig:2011-05-25}
\end{figure}

One transit of TrES-1 b, which was observed by the MicroObservatory Donald telescope on May 25, 2011, is shown as an example in Figure \ref{fig:2011-05-25}. It is fit by EXOTIC as described above. Each gray point represents the relative flux measured in one transit image. The blue points represent the moving average of the relative flux. The red curve represents the best-fit light curve for the data, where the transit begins and ends at a relative flux of $1.00$. The fitted mid-transit time is $2455707.8264 \pm 0.0014$ d, which is 0.81 minutes different from the expected mid-transit time for that epoch. The expected mid-transit time is calculated by finding the closest calculated transit midpoint time to the observed transit midpoint using \textcolor{black}{equation \eqref{midpt-keplerian-motion}}. \textcolor{black}{We used $3.03006973\pm 0.00000018$ d as the period and $2455016.96994 \text{ BJD}_\text{TDB} \pm 0.00007$ d as the initial mid-transit time \citep{Bonomo}.} The expected mid-transit time overlaps the uncertainty for the fitted mid-transit time. The RMS scatter of the residuals is $0.57\%$.

Capturing a complete exoplanet transit, especially with a small telescope such as those at the MicroObservatory, necessitates the perfect alignment of several factors.  Good seeing, lack of passing clouds, good tracking on the telescope while exposing, and minimal drift of the field of view between exposures all must conspire to capture clear images before, during, and after the target's occultation by its planet. Out of the 83 MicroObservatory transits in this study, 46 gave clear transit curves. These 46 transits were used for analysis.

\begin{figure}
    \centering
    \includegraphics[width=0.45\textwidth]{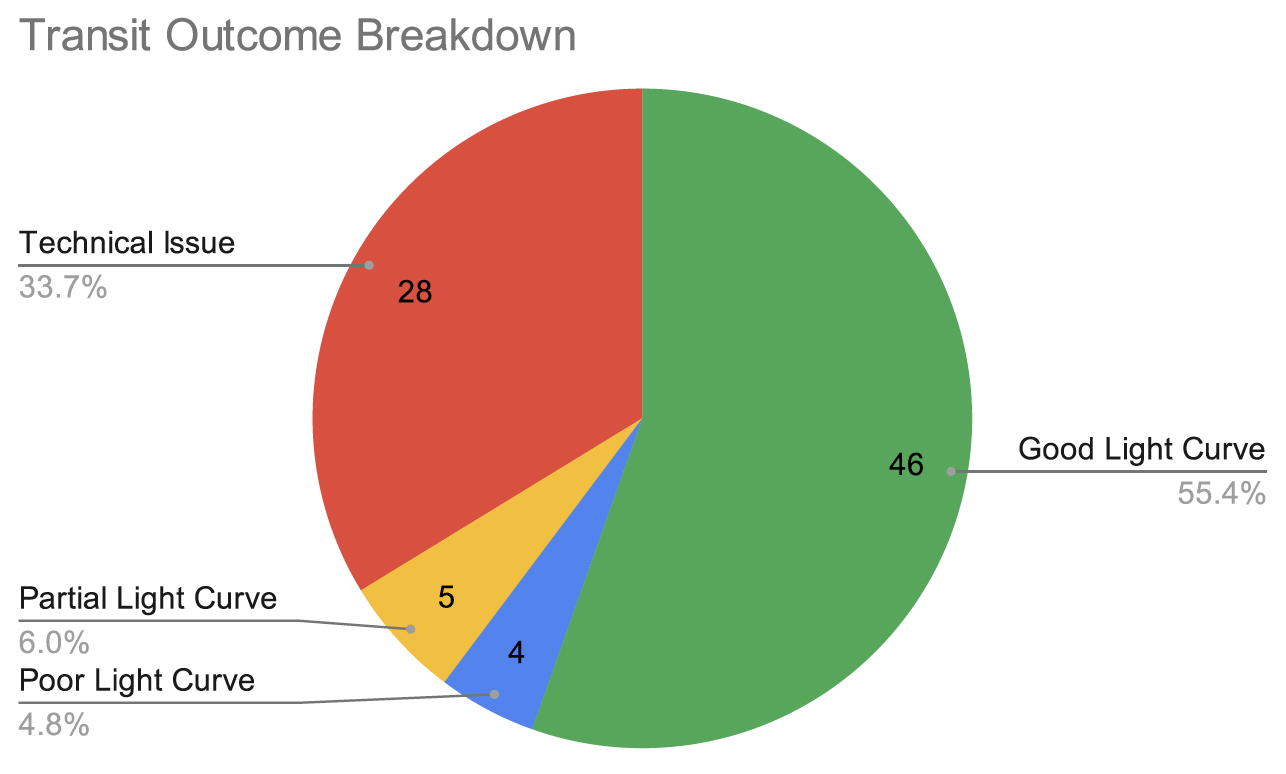}

    \caption{Breakdown of transit outcomes by light curve and image quality.}
    \label{fig:transit breakdown}
\end{figure}

The breakdown of the results from processing the set of images provided for each transit is shown in Figure \ref{fig:transit breakdown}. ``Good light curves'' were the light curves with a clear and complete transit (more precisely, there are observations both before the fitted transit begins and after it ends; the relative flux in the fitted light curve did not pass below 0.95; and scatter in residuals is not above 0.03). ``Poor light curves'' were those that appeared to show no clear transit even though the weather appeared normal in the images (more precisely, the relative flux in the fitted light curve did not pass below 0.95 or the scatter in residuals is above 0.03). ``Partial light curves'' were the light curves with clear transits that got cut off at either end (more precisely, the same criteria as for ``Good light curves'' are satisfied, except the the fitted transit either begins before the first observation or ends after the last observation). ``Technical issues'' was used to indicate that a transit produced an error while being processed or the images were unclear due to the weather or the telescopes (this could be detected by a variety of factors, but would be visible, for instance, through entirely washed out images or the target star leaving the field of view). Examples for each light curve designation are shown in Figure \ref{fig: light curves}. The light curves from each good transit are shown in the appendix in Figure \ref{fig:transit pics}.

\begin{figure}
    \centering
    \includegraphics[width=0.47\textwidth]{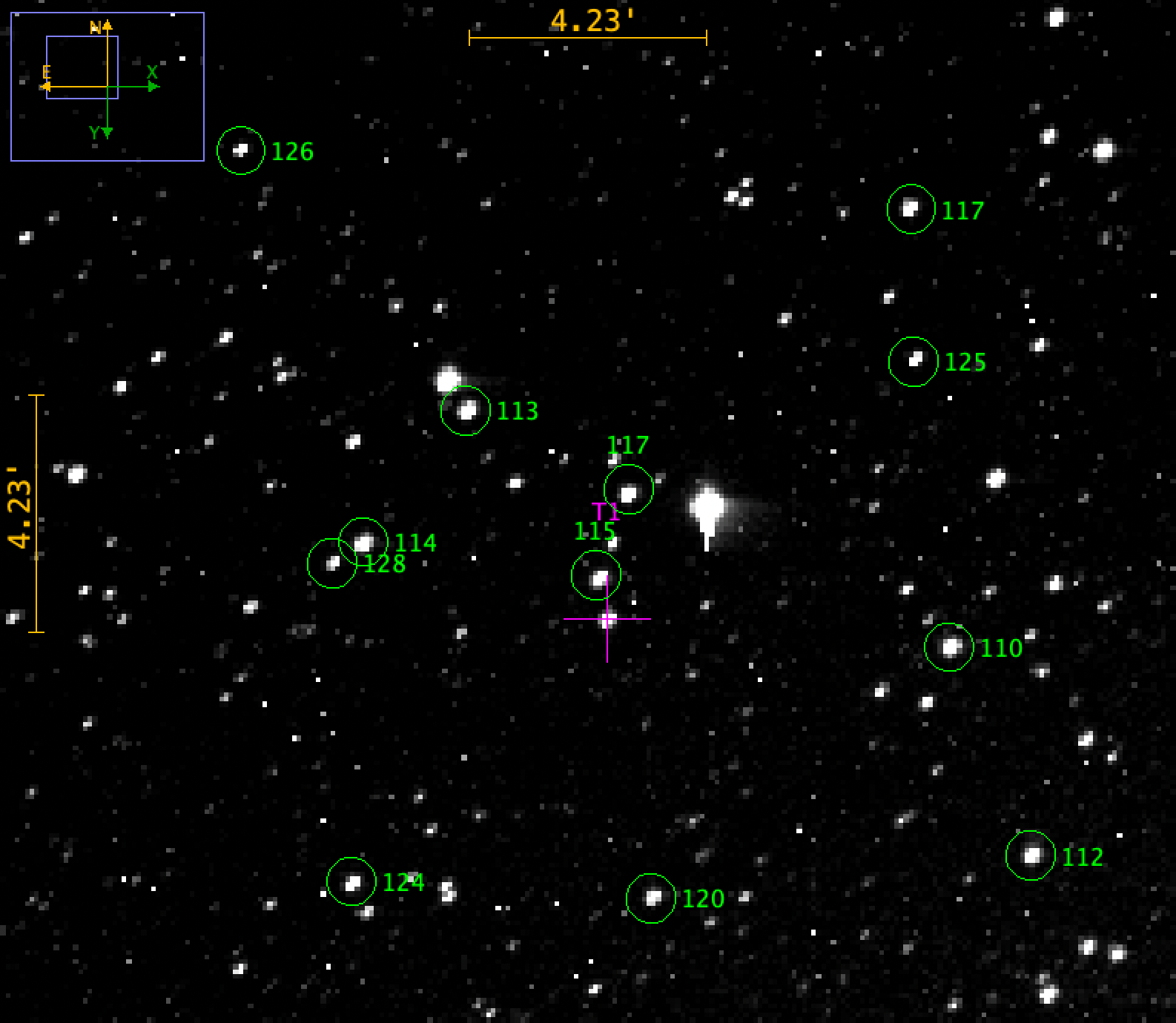}
    \caption{Star field of TrES-1 b with target labeled with ``T1'' and comparison stars labeled in green with their magnitudes.}
    \label{fig:starfield}
\end{figure}

For some of the transits that did not initially yield a good light curve, removing images that displayed poor weather or manually selecting different comparison stars shown in Figure \ref{fig:starfield} in sometimes helped.
Proximity of the comparison star to the target was also beneficial because light from both stars crossed a similar amount of Earth's atmosphere (``airmass'') on the way to the telescope, making their measured fluxes more comparable. Finally, the scatter in residuals of the light curves was minimized where the dark images were taken closer in date to the science images.

\begin{table*}
\centering
\caption{Transit midpoints and O-C values for Fig. \ref{fig:O-C MOBS} from the MicroObservatory and the Exoplanet Transit Database. The epoch and O-C values are calculated based on the period from the NASA Exoplanet Archive and the time of the first MicroObservatory observation.}
\hspace{-8em}
\resizebox{1.1\textwidth}{!}{%
    \begin{tabular}
    {| m{2cm} | m{3cm} | m{2.5cm} | m{2.5cm} | m{3.3cm} | m{2.5cm} | m{2.5cm} | m{2.5cm} | m{2.5cm}}
    \hline
    Date & Midpoint (BJD$_\text{TDB}$) & Midpoint Uncertainty \newline (BJD$_\text{TDB}$) & Epoch & O-C (min) & O-C Uncertainty (min) & Residual RMS Scatter & Source\\
    \hline
2004-09-20 & 2453268.62204 & 0.00073  & 0    & 3.36  & 1.05 & --    & ETD  \\ \hline
2004-09-23 & 2453271.65062 & 0.00120  & 1    & 1.22  & 1.73 & --    & ETD  \\ \hline
2005-04-26 & 2453486.78631 & 0.00110  & 72   & 2.28  & 1.58 & --    & ETD  \\ \hline
2006-04-30 & 2453856.45334 & 0.00050  & 194  & 0.15  & 0.72 & --    & ETD  \\ \hline
2007-09-03 & 2454347.32609 & 0.00028  & 356  & 2.25  & 0.41 & --    & ETD  \\ \hline
2007-09-06 & 2454350.35572 & 0.00036  & 357  & 1.61  & 0.52 & --    & ETD  \\ \hline
2009-05-16 & 2454968.48985 & 0.00028  & 561  & 1.48  & 0.41 & --    & ETD  \\ \hline
2009-06-16 & 2454998.79989 & 0.00160  & 571  & 14.93 & 2.31 & --    & ETD  \\ \hline
2009-08-18 & 2455062.42297 & 0.00060  & 592  & 2.86  & 0.87 & --    & ETD  \\ \hline
2010-04-18 & 2455304.82663 & 0.00084  & 672  & 0.09  & 1.22 & --    & ETD  \\ \hline
2010-06-20 & 2455368.46043 & 0.00068  & 693  & 3.46  & 0.99 & --    & ETD  \\ \hline
2010-10-19 & 2455489.6628  & 0.0013   & 733  & 2.86  & 1.88 & 0.007 & MOBS \\ \hline
2011-03-02 & 2455622.9834  & 0.0015   & 777  & -0.70 & 2.17 & 0.006 & MOBS \\ \hline
2011-04-19 & 2455671.46481 & 0.00090  & 793  & -0.28 & 1.31 & --    & ETD  \\ \hline
2011-04-29 & 2455680.55547 & 0.00064  & 796  & 0.37  & 0.93 & --    & ETD  \\ \hline
2011-05-19 & 2455701.7653  & 0.0016   & 803  & -0.57 & 2.31 & 0.008 & MOBS \\ \hline
2011-05-22 & 2455704.7959  & 0.0019   & 804  & 0.19  & 2.74 & 0.008 & MOBS \\ \hline
2011-05-25 & 2455707.8264  & 0.0014   & 805  & 0.81  & 2.02 & 0.005 & MOBS \\ \hline
2011-06-01 & 2455713.8859  & 0.0021   & 807  & -0.11 & 3.03 & 0.012 & MOBS \\ \hline
2011-06-04 & 2455716.9162  & 0.0019   & 808  & 0.22  & 2.74 & 0.007 & MOBS \\ \hline
2011-07-16 & 2455759.33911 & 0.00049  & 822  & 3.00  & 0.72 & --    & ETD  \\ \hline
2011-07-19 & 2455762.36788 & 0.00037  & 823  & 1.13  & 0.56 & --    & ETD  \\ \hline
2011-07-22 & 2455765.39965 & 0.00040  & 824  & 3.58  & 0.60 & --    & ETD  \\ \hline
2011-07-25 & 2455768.42996 & 0.00042  & 825  & 3.93  & 0.63 & --    & ETD  \\ \hline
2011-08-22 & 2455795.70127 & 0.00055  & 834  & 4.91  & 0.81 & --    & ETD  \\ \hline
2011-08-28 & 2455801.75817 & 0.00040  & 836  & 0.24  & 0.60 & --    & ETD  \\ \hline
2011-11-21 & 2455886.59868 & 0.00048  & 864  & -1.83 & 0.71 & --    & ETD  \\ \hline
2012-03-30 & 2456016.8927  & 0.0022   & 907  & -0.36 & 3.17 & 0.010 & MOBS \\ \hline
2012-04-02 & 2456019.9238  & 0.0020   & 908  & 1.13  & 2.89 & 0.009 & MOBS \\ \hline
2012-05-23 & 2456071.43362 & 0.00070  & 925  & -0.84 & 1.03 & --    & ETD  \\ \hline
2012-05-26 & 2456074.46528 & 0.00112  & 926  & 1.45  & 1.62 & --    & ETD  \\ \hline
2012-06-17 & 2456095.67860 & 0.00087  & 933  & 5.52  & 1.27 & --    & ETD  \\ \hline
2012-06-19 & 2456098.7020  & 0.0016   & 934  & -4.08 & 2.31 & 0.010 & MOBS \\ \hline
2012-06-22 & 2456101.7363  & 0.0022   & 935  & 2.02  & 3.17 & 0.011 & MOBS \\ \hline
2012-06-29 & 2456107.79744 & 0.00032  & 937  & 3.45  & 0.50 & --    & ETD  \\ \hline
2012-07-04 & 2456113.8545  & 0.0025   & 969  & 2.05  & 3.75 & 0.026 & MOBS \\ \hline
2012-09-28 & 2456198.69781 & 0.00056  & 967  & 0.97  & 0.83 & --    & ETD  \\ \hline
2012-10-03 & 2456204.7570  & 0.0020   & 1038 & -0.39 & 3.61 & 0.011 & MOBS \\ \hline
2013-05-01 & 2456413.8300  & 0.0013   & 1040 & -3.00 & 2.89 & 0.010 & MOBS \\ \hline
2013-05-07 & 2456419.8891  & 0.0017   & 1041 & -4.50 & 1.88 & 0.006 & MOBS \\ \hline
2013-05-10 & 2456422.9235  & 0.0016   & 1072 & 1.74  & 2.46 & 0.006 & MOBS \\ \hline
2013-07-03 & 2456477.46917 & 0.00042  & 1059 & 8.09  & 0.65 & --    & ETD  \\ \hline
2013-07-06 & 2456480.49645 & 0.00035  & 1060 & 4.08  & 0.55 & --    & ETD  \\ \hline
2013-07-31 & 2456504.73647 & 0.00030  & 1068 & 3.30  & 0.49 & --    & ETD  \\ \hline
2013-08-11 & 2456516.8567  & 0.0027   & 1100 & 3.24  & 2.32 & 0.006 & MOBS \\ \hline
2013-11-04 & 2456601.6937  & 0.0009   & 1168 & -3.90 & 3.90 & 0.014 & MOBS \\ \hline
2014-05-06 & 2456783.50156 & 0.00055  & 1160 & 1.39  & 0.83 & --    & ETD  \\ \hline
2014-05-29 & 2456807.7358  & 0.0016   & 1169 & -7.73 & 1.32 & 0.010 & MOBS \\ \hline
2014-06-01 & 2456810.7747  & 0.0011   & 1170 & 5.02  & 2.32 & 0.007 & MOBS \\ \hline
2014-06-04 & 2456813.7968  & 0.0014   & 1171 & -6.46 & 1.60 & 0.006 & MOBS \\ \hline

\end{tabular}%
}
    
    \label{tab:EXOTIC data}
\end{table*}

\setcounter{table}{0}

\begin{table*}
\centering
\caption{Transit midpoints and O-C values for Fig. \ref{fig:O-C MOBS} from the MicroObservatory and the Exoplanet Transit Database. The epoch and O-C values are calculated based on the period from the NASA Exoplanet Archive and the time of the first MicroObservatory observation.}
\hspace{-8em}
\resizebox{1.1\textwidth}{!}{%
    \begin{tabular}
    {| m{2cm} | m{3cm} | m{2.5cm} | m{2.5cm} | m{3.3cm} | m{2.5cm} | m{2.5cm} | m{2.5cm} | m{2.5cm}}
    \hline
    Date & Midpoint (BJD$_\text{TDB}$) & Midpoint Uncertainty \newline (BJD$_\text{TDB}$) & Epoch & O-C (min) & O-C Uncertainty (min) & Residual RMS Scatter & Source\\
    \hline
    2014-06-07 & 2456816.8320  & 0.0017   & 1304 & 0.93  & 2.03 & 0.005 & MOBS \\ \hline
2014-08-10 & 2456880.46568 & 0.00035  & 1192 & 4.11  & 0.57 & --    & ETD  \\ \hline
2014-08-13 & 2456883.49504 & 0.00064  & 1193 & 3.09  & 0.96 & --    & ETD  \\ \hline
2014-08-17 & 2456886.52910 & 0.00136  & 1194 & 8.84  & 1.98 & --    & ETD  \\ \hline
2014-08-29 & 2456898.64521 & 0.00049  & 1198 & 2.83  & 0.75 & --    & ETD  \\ \hline
2015-04-10 & 2457122.86795 & 0.00102  & 1272 & -0.65 & 1.50 & --    & ETD  \\ \hline
2015-07-15 & 2457219.8340  & 0.0017   & 1332 & 4.85  & 2.47 & 0.007 & MOBS \\ \hline
2015-10-08 & 2457304.6669  & 0.0035   & 1435 & -8.18 & 2.47 & 0.009 & MOBS \\ \hline
2016-04-20 & 2457498.59704 & 0.00047  & 1396 & -0.01 & 0.75 & --    & ETD  \\ \hline
2016-07-16 & 2457586.47109 & 0.00038  & 1425 & 2.91  & 0.63 & --    & ETD  \\ \hline
2016-07-19 & 2457589.49869 & 0.00107  & 1426 & -0.65 & 1.57 & --    & ETD  \\ \hline
2016-08-21 & 2457622.8288  & 0.0042   & 1437 & -1.59 & 6.06 & 0.020 & MOBS \\ \hline
2016-11-08 & 2457701.6111  & 0.0043   & 1463 & -0.89 & 6.20 & 0.017 & MOBS \\ \hline
2016-11-11 & 2457704.6438  & 0.0033   & 1464 & 2.90  & 4.76 & 0.015 & MOBS \\ \hline
2017-05-28 & 2457901.59874 & 0.00077  & 1529 & 3.48  & 1.16 & --    & ETD  \\ \hline
2017-06-11 & 2457916.7425  & 0.0030   & 1534 & -6.00 & 4.33 & 0.014 & MOBS \\ \hline
2017-06-14 & 2457919.7724  & 0.0027   & 1535 & -6.25 & 3.90 & 0.011 & MOBS \\ \hline
2017-08-23 & 2457989.47123 & 0.00046  & 1558 & 4.15  & 0.75 & --    & ETD  \\ \hline
2018-04-19 & 2458228.8500  & 0.0018   & 1637 & 8.85  & 2.62 & 0.010 & MOBS \\ \hline
2018-04-22 & 2458231.8734  & 0.0023   & 1638 & -0.75 & 3.33 & 0.009 & MOBS \\ \hline
2018-06-16 & 2458286.41955 & 0.00057  & 1656 & 6.29  & 0.90 & --    & ETD  \\ \hline
2018-06-19 & 2458289.44708 & 0.00080  & 1657 & 2.64  & 1.21 & --    & ETD  \\ \hline
2018-06-22 & 2458292.47855 & 0.00037  & 1658 & 4.65  & 0.66 & --    & ETD  \\ \hline
2018-07-19 & 2458319.7450  & 0.0029   & 1667 & -1.36 & 4.19 & 0.015 & MOBS \\ \hline
2018-08-04 & 2458334.8993  & 0.0029   & 1672 & 4.33  & 4.19 & 0.012 & MOBS \\ \hline
2018-09-27 & 2458389.43702 & 0.00111  & 1690 & -0.76 & 1.65 & --    & ETD  \\ \hline
2018-10-18 & 2458410.6461  & 0.0019   & 1697 & -2.79 & 2.76 & 0.012 & MOBS \\ \hline
2018-10-21 & 2458413.6715  & 0.0023   & 1698 & -9.51 & 3.34 & 0.015 & MOBS \\ \hline
2019-05-21 & 2458625.7833  & 0.0024   & 1768 & 0.45  & 3.48 & 0.022 & MOBS \\ \hline
2019-05-27 & 2458631.8395  & 0.0029   & 1770 & -5.22 & 4.20 & 0.009 & MOBS \\ \hline
2019-06-03 & 2458637.9001  & 0.0021   & 1772 & -4.56 & 3.05 & 0.008 & MOBS \\ \hline
2019-07-24 & 2458689.41747 & 0.00039  & 1789 & 4.34  & 0.70 & --    & ETD  \\ \hline
2019-08-03 & 2458698.50695 & 0.00056  & 1792 & 3.29  & 0.91 & --    & ETD  \\ \hline
2019-08-06 & 2458701.53762 & 0.00056  & 1793 & 4.16  & 0.91 & --    & ETD  \\ \hline
2019-08-09 & 2458704.56817 & 0.00038 & 1794 & 4.85  & 0.69 & --    & ETD  \\ \hline
2019-08-20 & 2458716.6835  & 0.0014   & 1798 & -2.27 & 2.06 & 0.005 & MOBS \\ \hline
2019-08-26 & 2458722.7471  & 0.0013   & 1800 & 2.71  & 1.92 & 0.006 & MOBS \\ \hline
2019-09-01 & 2458728.8062  & 0.0021   & 1802 & 1.21  & 3.05 & 0.009 & MOBS \\ \hline
2019-09-04 & 2458731.8356  & 0.0023   & 1803 & 0.25  & 3.34 & 0.009 & MOBS \\ \hline
2020-06-28 & 2459028.7843  & 0.0030   & 1901 & 2.94  & 4.34 & 0.009 & MOBS \\ \hline
2020-07-04 & 2459034.8377  & 0.0015   & 1903 & -6.77 & 2.21 & 0.008 & MOBS \\ \hline
2020-09-27 & 2459119.6802  & 0.0016   & 1931 & -5.98 & 2.35 & 0.008 & MOBS \\ \hline
2020-09-30 & 2459122.7205  & 0.0027   & 1932 & 8.75  & 3.91 & 0.012 & MOBS \\ \hline

\end{tabular}%
}
    \label{tab:EXOTIC data}
\end{table*}

\begin{figure*}
 \centering
  \includegraphics[width=0.4\textwidth]{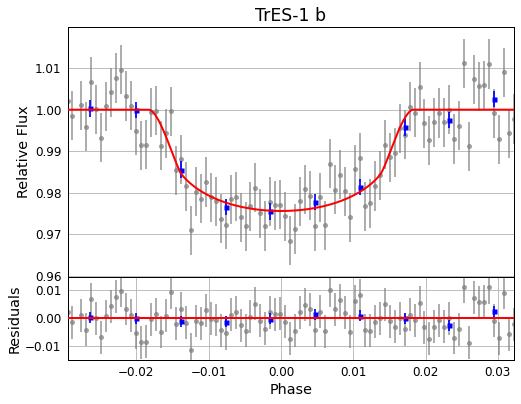} \includegraphics[width=0.4\textwidth]{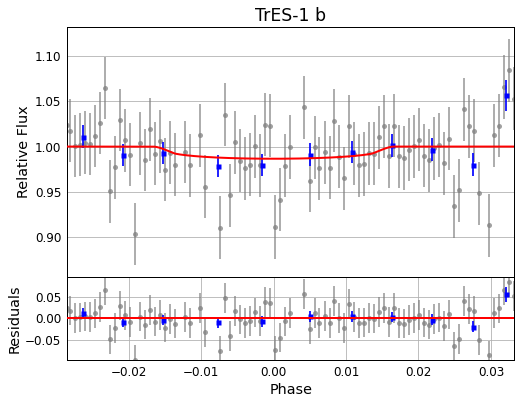} \includegraphics[width=0.4\textwidth]{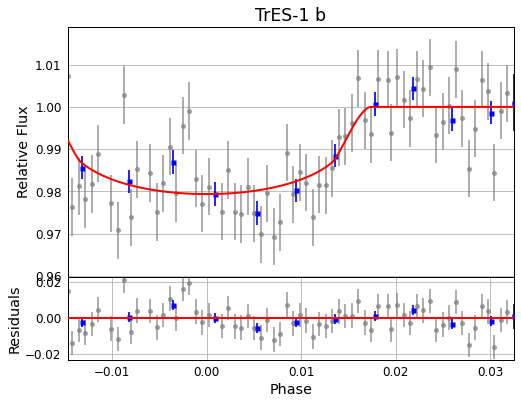}
  \includegraphics[width=0.4\textwidth]{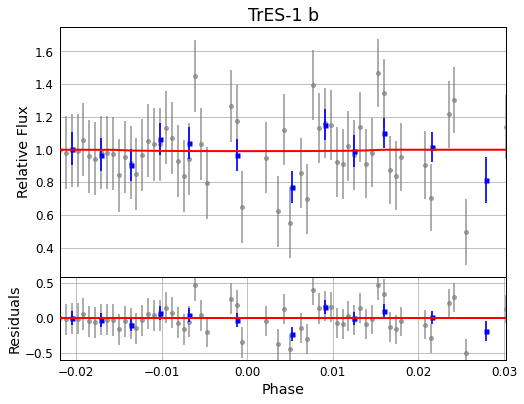}
  \caption{Top left: ``good light curve'' from images taken on May 25, 2011. Top right: ``poor light curve'' from images taken on June 25, 2020. Bottom left: ``partial light curve'' from images taken on March 26, 2012. Bottom right: ``technical issue'' from images taken on July 10, 2020.}
   \label{fig: light curves}
\end{figure*}

\begin{figure*}
    \centering
    \includegraphics[width=0.525\textwidth]{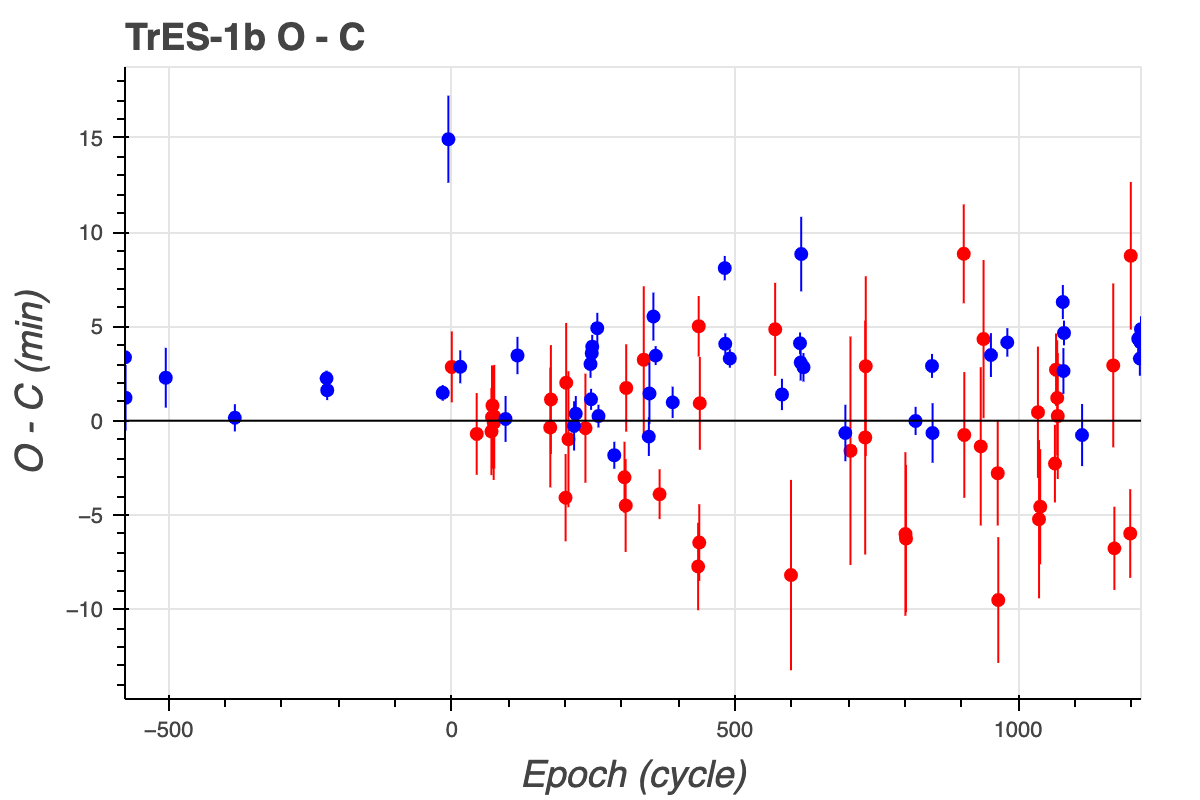}
    \caption{O-C plot from MicroObservatory data (plotted in red) along with archival data (plotted in blue) compared to the predicted value based on the period from the NASA Exoplanet Archive and the time of the first MicroObservatory observation.}
    \label{fig:O-C MOBS}
\end{figure*}


\section{Transit Timing Variations} 
\textcolor{black}{In this section the procedures for calculating and analyzing TTVs are examined.} \textcolor{black}{First, the observed transits are fitted to attain the transit midpoints and other parameters.} Then, for a planet in Keplerian motion, the transit midpoint time can be described by the equation
\begin{equation}
   T = T_0 + E\cdot P
  \label{midpt-keplerian-motion}
\end{equation}
where $T$ is a given transit midpoint, $E$ is the epoch number (indexed such that $T_0$ is the midpoint of the transit at epoch $0$), and $P$ is the period of the planet \citep{corteszuleta2020tramos}. The transit midpoint can be predicted based on how the transit midpoint would change over time given that the motion is periodic. Based on this equation and using published values for $T_0$ and $P$, it is possible to find observed minus calculated (O-C) values for each analyzed observation by subtracting the observed transit midpoint from the closest calculated time. Using equation $(3)$ in \cite{ETS} for calculated mid-transit time uncertainty and using \textcolor{black}{equation \eqref{midpt-keplerian-motion}}, the uncertainty of the O-C value can be calculated to be
\small
$$\sqrt{\Delta T_{obs}^2 + E^2\cdot\Delta P^2 + 2\cdot E\cdot\Delta P\cdot\Delta T_0 + \Delta T_0^2}.$$
\normalsize
Next, comparing the observed transit midpoints to the prediction based on the calculated O-C values, the extent to which other planets in the system might be affecting the system's period can be assessed. \textcolor{black}{From here, the mass of possible planets can be bounded based on perturbations these planets would cause on the orbit of the observed planet, and Lomb-Scargle period analysis can be used to check for a periodic trend among observed residuals \citep{VanderPlas_2018}.}

The transit midpoints calculated by EXOTIC and the midpoints from archival data in the Exoplanet Transit Database (ETD)\footnote{http://var2.astro.cz/ETD/} were compared to the expected midpoints based on the listings in the NASA Exoplanet Archive and the chronologically first observations from the MicroObservatory (MOBS) and ETD. In particular, the period used was $3.03006973\pm 0.00000018$ d \citep{Bonomo}. The initial mid-transit times used were $2453268.62204\pm 0.00073$ BJD$_{\text{TDB}}$  and $2455489.6628 \pm 0.0013$ BJD$_{\text{TDB}}$ from ETD and MOBS, respectively.


Figure \ref{fig:O-C MOBS} shows a plot of the observed transit midpoints versus the expected transit midpoints using the chronologically first observation and reference time from the MicroObservatory data. The processed data that were used to create Figure \ref{fig:O-C MOBS} are seen in Table \ref{tab:EXOTIC data}. Using 30 degrees of freedom, the reduced $\chi^2$ values of the data in the plot compared to the line $\text{O-C} = 0$ was $1.06$. This low $\chi^2$ value suggests that the data are consistent with the calculated transit midpoints.


Based on the processed data and the archival data, an updated ephemeris was fit to the data. The result of the fit was $T = (2455489.66026 \text{ BJD}_\text{TDB} \pm 0.00044 \text{ d}) + (3.0300689 \pm 0.0000007)\text{ d}\cdot E$, where $T$ is the transit time in BJD$_\text{TDB}$ and $E$ is the epoch number after $2458731.83$ BJD$_\text{TDB}$. The resulting period, $3.0300689 \pm 0.0000007$ d, is consistent with the archival time from the NASA Exoplanet Archive of $3.03006973\pm 0.00000018$ d as the intervals overlap \citep{Bonomo}.

\begin{figure}[h]
    \centering
    \includegraphics[width=0.45\textwidth]{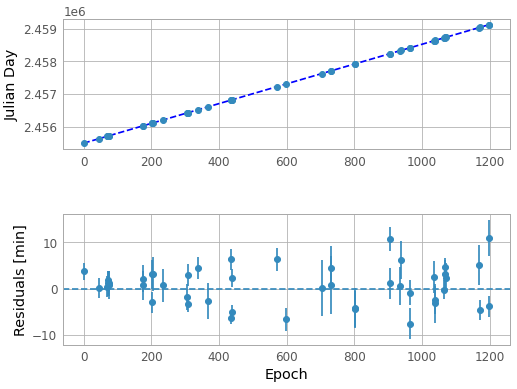}
    \includegraphics[width=0.35\textwidth]{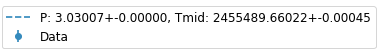}
    \caption{Graph of archival and processed transit midpoints along with the fitted linear ephemeris.}
    \label{fig:fittedephemeris}
\end{figure}

The graph of the fitted ephemeris and residuals is shown in Figure \ref{fig:fittedephemeris}.

As shown in Figure \ref{fig:O-C MOBS}, the uncertainties of the transit midpoints from archival data are much smaller than the uncertainties from the data we reduced using EXOTIC. While EXOTIC uses a Markov Chain Monte Carlo method to compute the uncertainties, the uncertainties from the ETD are calculated from Poisson statistics and readout noise, and they are lower because they do not factor in red noise \citep{PODDANY2010297}.

\section{Lomb-Scargle Analysis}
After the curves are fitted through EXOTIC, the Lomb-Scargle method was used to search for a periodicity in the O-C values.

The Lomb-Scargle algorithm for period analysis is suitable for unevenly sampled time-series data \citep{VanderPlas_2018}. It is used to determine if there are significant periodic trends in transit data rather than simply random Gaussian noise.




The Lomb-Scargle periodogram produced by this procedure is modified from the classical periodogram, which is based directly on the Fourier transform. The classical periodogram does not work well for analyzing non-uniform data because the non-uniformity in the data translates to a ``noisy'' Fourier transform with irregular frequency peaks \citep{VanderPlas_2018}. This noisy translation does not happen with the Lomb-Scargle modification of the classical periodogram, so we use the Lomb-Scargle procedure for a more accurate test of periodicity. We performed the procedure using the \texttt{astropy.timeseries.LombScargle} module.

\begin{figure*}
    \centering
    \includegraphics[width=\textwidth]{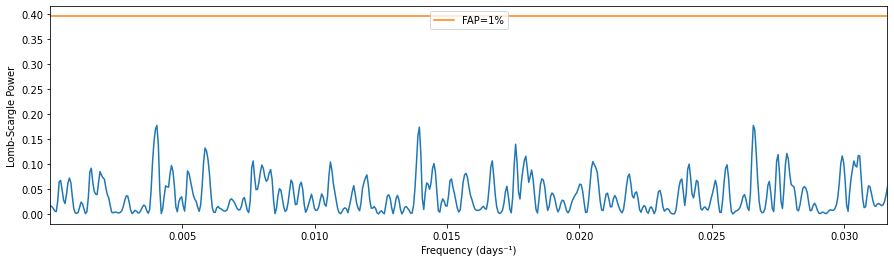}
    \caption{Lomb-Scargle periodogram produced from O-C values with data from 46 of the transits we reduced.}
    \label{fig:LS}
\end{figure*}
\begin{figure*}
    \centering
    \includegraphics[width=\textwidth]{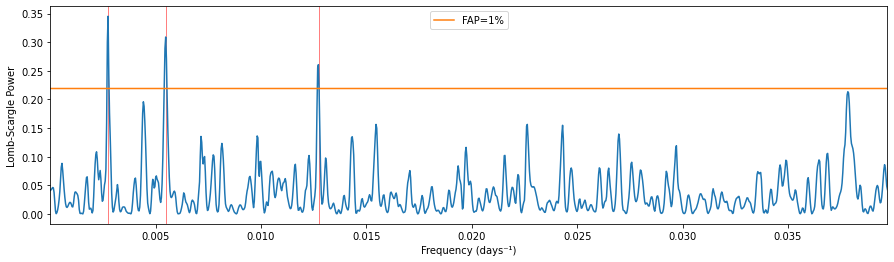}
    \caption{Lomb-Scargle periodogram produced from O-C values with data from 46 of the transits we reduced and 47 transits from the Exoplanet Transit Database. The three highest peaks with frequencies of 0.00275, 0.00548, and 0.01276 cycles per day (in decreasing order of power) are highlighted in red.}
    \label{fig:LScombined}
\end{figure*}

The periodograms in Figures \ref{fig:LS} and \ref{fig:LScombined} produced using the Lomb-Scargle procedure were created from the MicroObservatory data of 46 TrES-1 b transits and the combined MicroObservatory and Exoplanet Transit Database data of 93 transits, respectively.

In Figure \ref{fig:LS}, the highest peak has a false-alarm probability of 0.98, suggesting that the data do not follow a periodic trend. The small number of observations being analyzed may also contribute to the high false-alarm probability as the presence or absence of a periodic signal is less likely to be accurately detected in a smaller sample size.

The graph for the combined data in Figure \ref{fig:LScombined} shows three significant peaks at frequencies of 0.00275, 0.00548, and 0.01276 cycles per day. These frequencies correspond to periods of 363.6, 182.4, and 78.4 days, respectively. Each of the three peaks has a false-alarm probability under 0.01, indicating that the data may follow a periodic trend. This finding requires further investigation to determine whether additional planets are present in the system.


\section{Future Work}

One interesting direction could be to look at cumulative and individual reduction of transits could be compared to see which yields lower uncertainty. Were the transits not all from the same telescope, it could be considered whether the transits could be combined without modification or how measurements could be adjusted or weighted. Weighting schemes could be considered based on quality of the night or similar factors for the same telescope as well.

Another possible direction would be to look into further automating the processing of all of the transits to speed up the process. This could involve automating classification of transits as ``good light curves,'' ``poor light curves,'' ``partial light curves,'' and ``technical issues'' to help determine which transits to use. This automation could allow for improvement over the image filtering currently included in EXOTIC, as some of the transits we analyzed required further manual removal of problematic images. It is also possible to look into using cloud computing services to parallelize the processing of transits, which would drastically reduce the amount of time and work it takes to process many transits.


\section{Conclusions}

Measurements and analysis of transit timing variations in TrES-1 b transits are presented in this paper. TTV analysis based on these transits suggests the possibility of the existence of another planet, which requires further investigation.
It is important to continue observing such transits to maintain ephemerides and provide targets for future transiting exoplanet atmospheric characterization missions \citep{ETS}. Our contribution of reduction of follow-up observations of TrES-1 b and analysis of TTVs has been detailed.

\section{Acknowledgments}

This publication makes use of the EXOTIC data reduction package from Exoplanet Watch, a citizen science project managed by NASA’s Jet Propulsion Laboratory on behalf of NASA’s Universe of Learning. This work is supported by NASA under award number NNX16AC65A to the Space Telescope Science Institute.

This research has made use of the data provided by the MicroObservatory telescope network in addition to the \texttt{astropy.io.fits} and \texttt{astropy.timeseries.LombScargle} 
Python modules.

This research has also incorporated historical data from the Exoplanet Transit Database. Special thanks to Yves Jongen, Vicenç Ferrando, Aleksandra Selezneva, Mario Morales, Rene Roy, Marc Bretton, Anaël Wünsche, Mohammed Talafha, Francisco Jiménez Alvarado, Dominika Ďurovčíková, Karol Petrík, Ramon Naves, David Molina, Sean Balkwill, Alexander, Jennifer Eastman, Steffen Shaigec of the Athabasca University Geophysical Observatory, Mickie Wiebe, Terry Youngman, Kevin B. Alton, Ferran Grau Horta, Mark Salisbury, Alessandro Marchini, Juanjo Gonzalez, Katie Iadanza, Thomas Balonek, Fernand Emering, Bradley Walter et al. of the Paul and Jane Meyer Observatory (PJMO), Viktoriia Krushevska and Yuliana Kuznyetsova of the Main Astronomical Observatory of the NAS of Ukraine, Maksim V. Andreev of the ICAMER Observatory of the NAS of Ukraine, Darryl Sergison, Martin Vrašťák, Luboš Brát, Manfred Raetz, Stan Shadick, Thomas Sauer, Radek Dřevěný, Tomáš Kalisch, Joe Garlitz, and Bruce Gary for providing these observations.

Further, we would like to thank the AAVSO observers with observer codes GELD, HTAA, HAMA, HLAC, MTRC, NCAA, OXEA, RAKB, SHAF, SJOR, AKV , URMA, KMUA, KADB, LGEC, VJCA, SNIC, KELA, FMAA, DJJB, JBEB, GBRC, and LDJC for uploading transits of TrES-1 b to the AAVSO database.

We would also like to thank Frank Sienkiewicz at the Center for Astrophysics $|$ Harvard \& Smithsonian for his work in maintaining the MicroObservatory telescopes and making the observations we relied on possible.


\pagebreak
\clearpage
\newpage

\bibliographystyle{plainnat}
\bibliography{Bibliography.bib}

\section{Appendix}

\begin{figure*}[htbp]
\subfloat[2010-10-19]{
\includegraphics[width=0.23\textwidth]{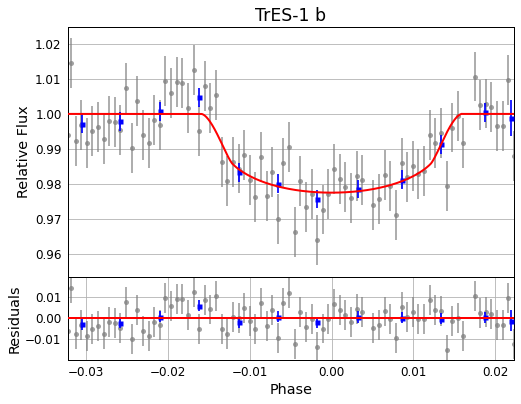}
}\subfloat[2011-03-02]{
\includegraphics[width=0.23\textwidth]{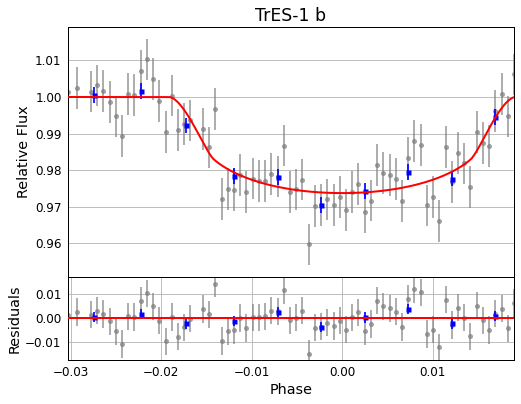}
}\subfloat[2011-05-19]{
\includegraphics[width=0.23\textwidth]{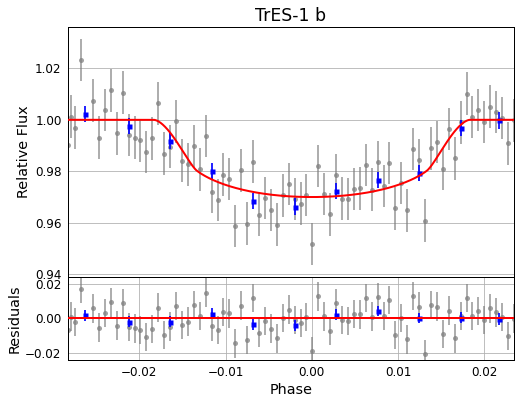}
}\subfloat[2011-05-22]{
\includegraphics[width=0.23\textwidth]{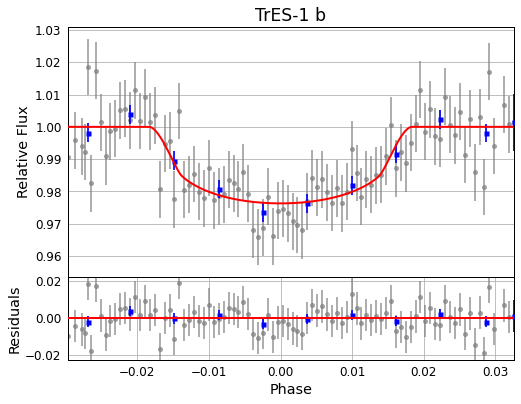}
}
\\
\subfloat[2011-05-25]{
\includegraphics[width=0.23\textwidth]{lightcurves/FinalLightCurve_TrES-1_b_2011-05-25.png}
}\subfloat[2011-06-01]{
\includegraphics[width=0.23\textwidth]{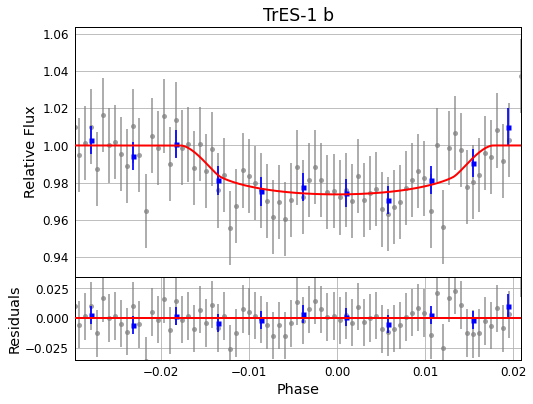}
}\subfloat[2011-06-04]{
\includegraphics[width=0.23\textwidth]{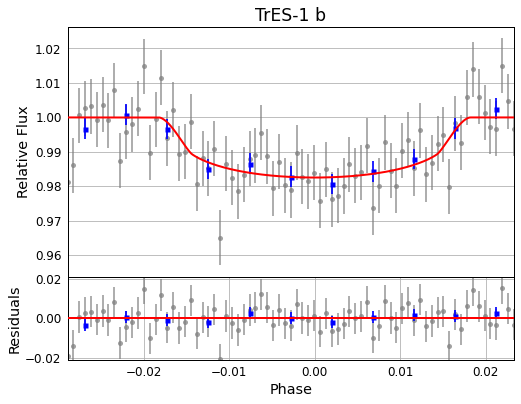}
}\subfloat[2012-03-30]{
\includegraphics[width=0.23\textwidth]{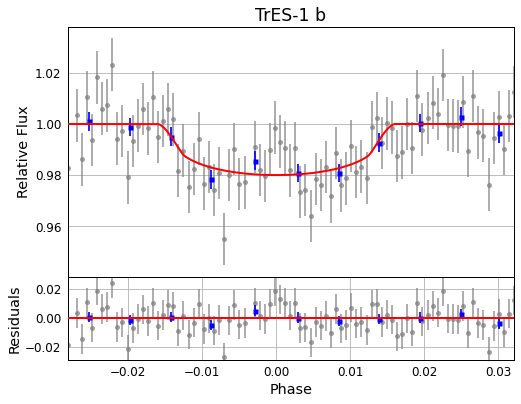}
}
\\
\subfloat[2012-04-02]{
\includegraphics[width=0.23\textwidth]{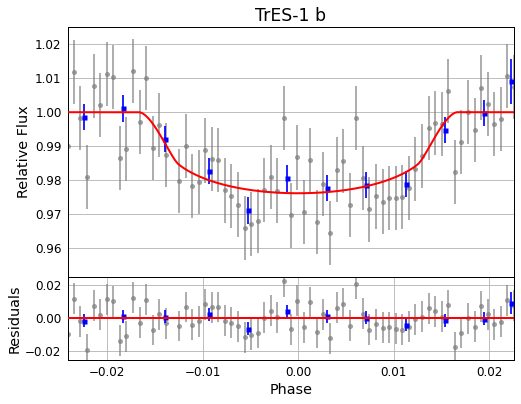}
}\subfloat[2012-06-19]{
\includegraphics[width=0.23\textwidth]{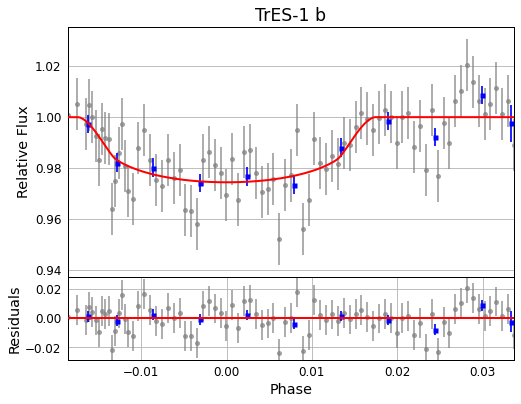}
}\subfloat[2012-06-22]{
\includegraphics[width=0.23\textwidth]{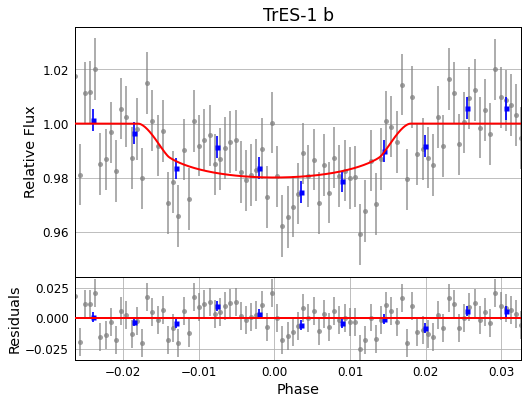}
}\subfloat[2012-07-04]{
\includegraphics[width=0.23\textwidth]{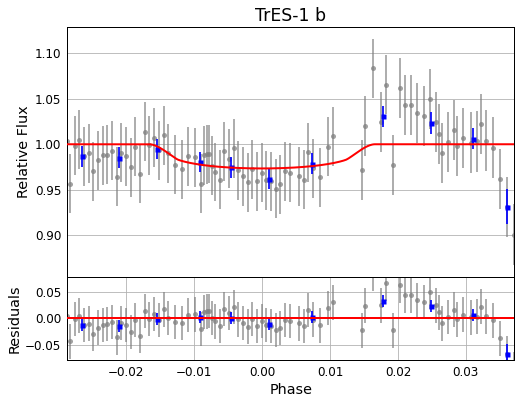}
}
\\
\subfloat[2012-10-03]{
\includegraphics[width=0.23\textwidth]{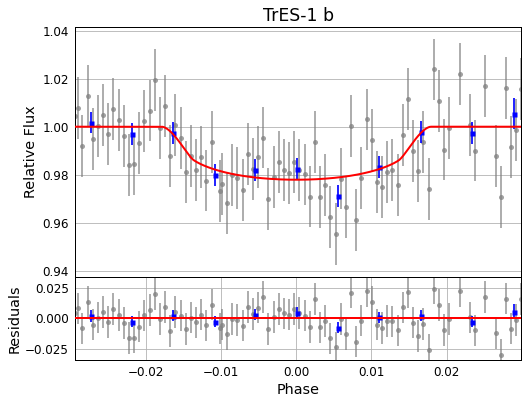}
}\subfloat[2013-05-01]{
\includegraphics[width=0.23\textwidth]{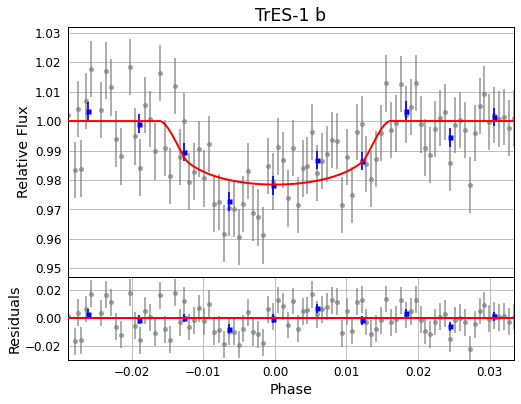}
}\subfloat[2013-05-07]{
\includegraphics[width=0.23\textwidth]{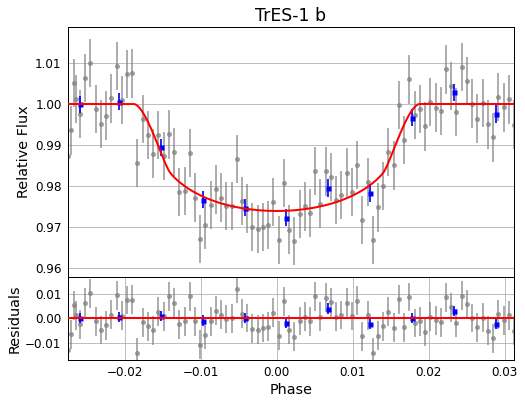}
}\subfloat[2013-05-10]{
\includegraphics[width=0.23\textwidth]{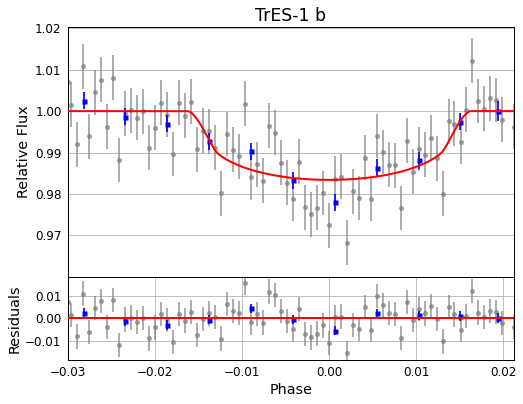}
}
\\
\subfloat[2013-08-11]{
\includegraphics[width=0.23\textwidth]{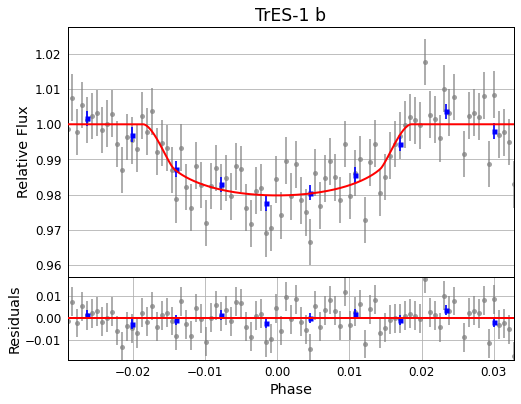}
}\subfloat[2013-11-04]{
\includegraphics[width=0.23\textwidth]{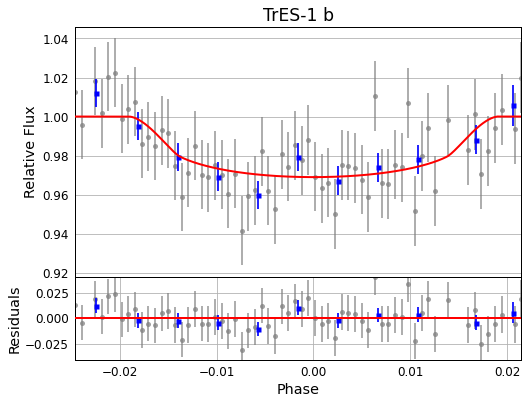}
}\subfloat[2014-05-29]{
\includegraphics[width=0.23\textwidth]{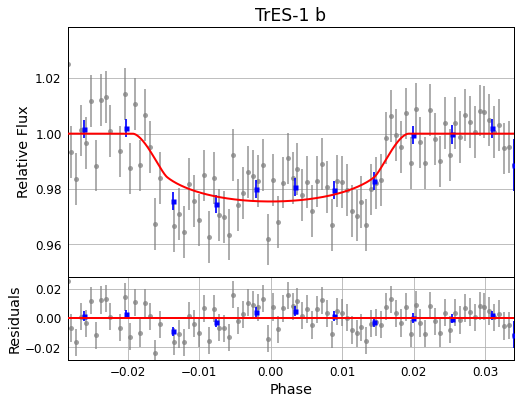}
}\subfloat[2014-06-01]{
\includegraphics[width=0.23\textwidth]{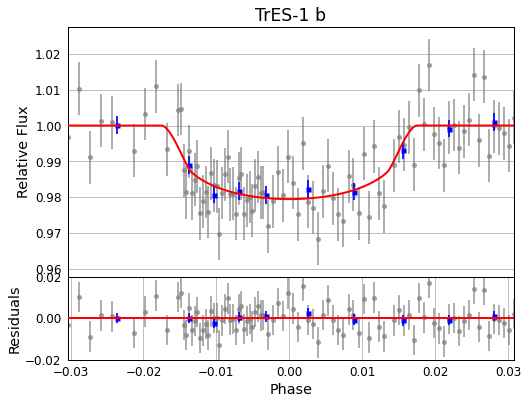}
}
\caption{Images of all good light curves. The date format is YYYY-MM-DD.}
\label{fig:transit pics}
\end{figure*}
\begin{figure*}[htbp]
\ContinuedFloat
\subfloat[2014-06-04]{
\includegraphics[width=0.23\textwidth]{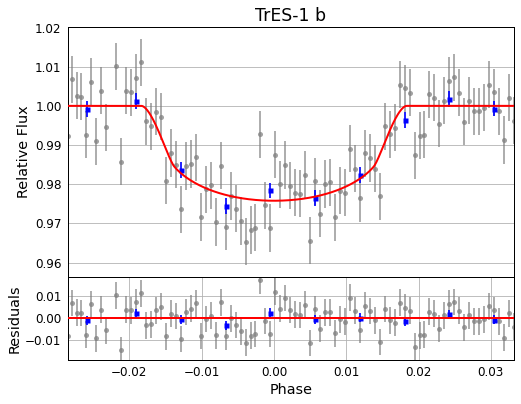}
}\subfloat[2014-06-07]{
\includegraphics[width=0.23\textwidth]{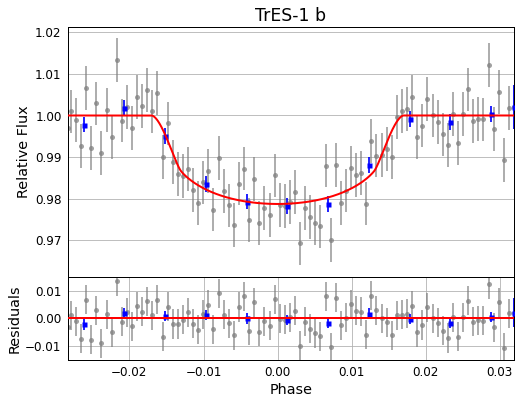}
}\subfloat[2015-07-15]{
\includegraphics[width=0.23\textwidth]{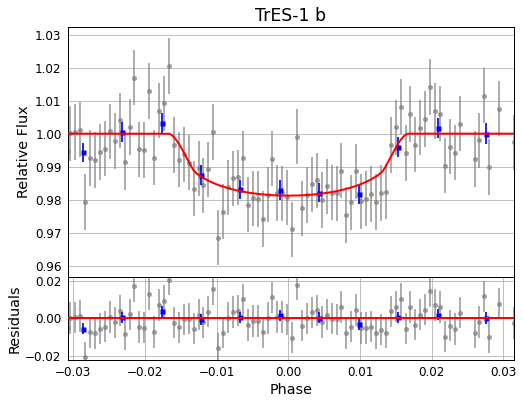}
}\subfloat[2015-10-08]{
\includegraphics[width=0.23\textwidth]{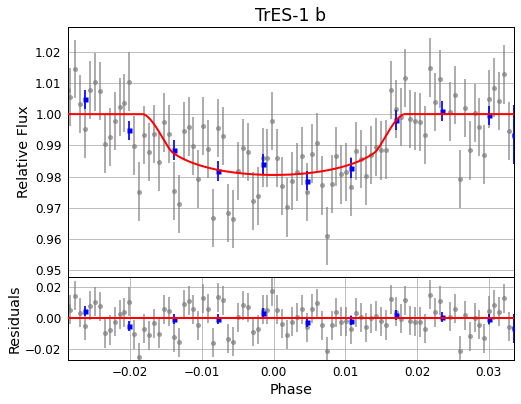}
}
\\
\subfloat[2016-08-21]{
\includegraphics[width=0.23\textwidth]{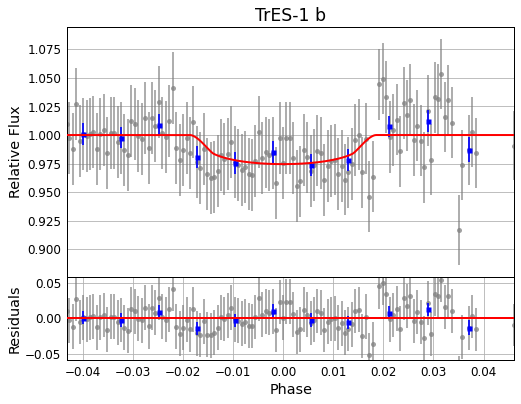}
}\subfloat[2016-11-08]{
\includegraphics[width=0.23\textwidth]{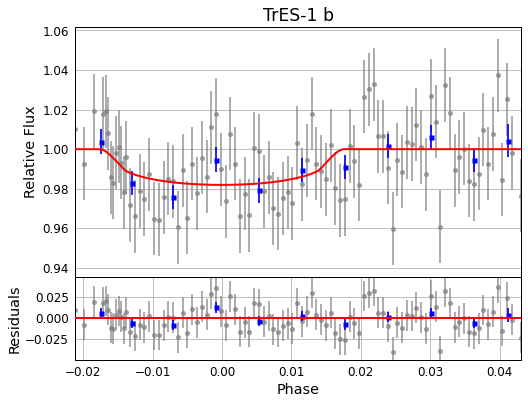}
}\subfloat[2016-11-11]{
\includegraphics[width=0.23\textwidth]{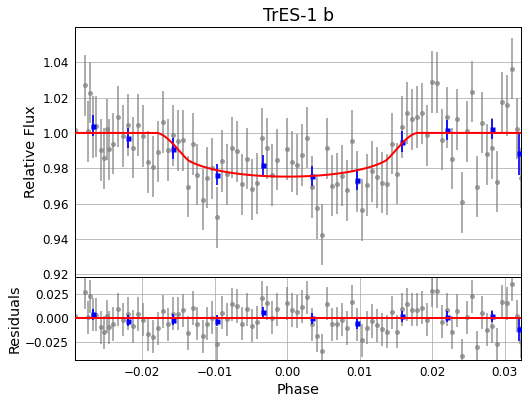}
}\subfloat[2017-06-11]{
\includegraphics[width=0.23\textwidth]{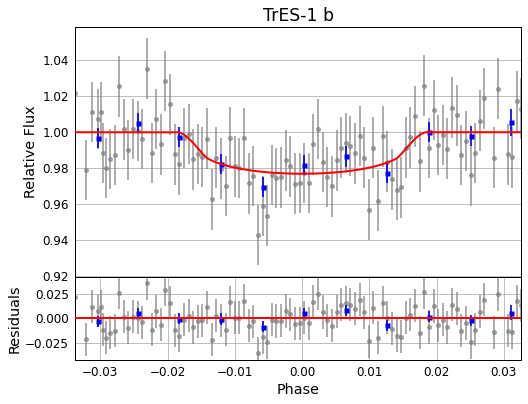}
}
\\
\subfloat[2017-06-14]{
\includegraphics[width=0.23\textwidth]{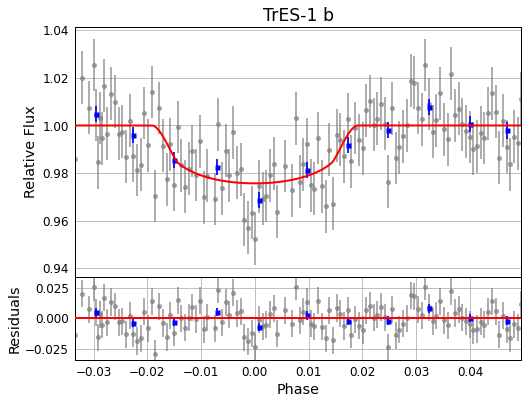}
}\subfloat[2018-04-19]{
\includegraphics[width=0.23\textwidth]{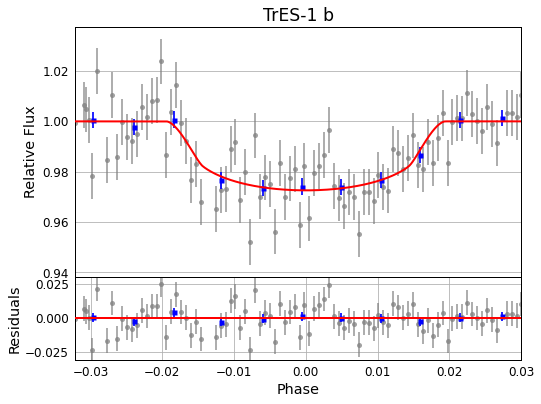}
}\subfloat[2018-04-22]{
\includegraphics[width=0.23\textwidth]{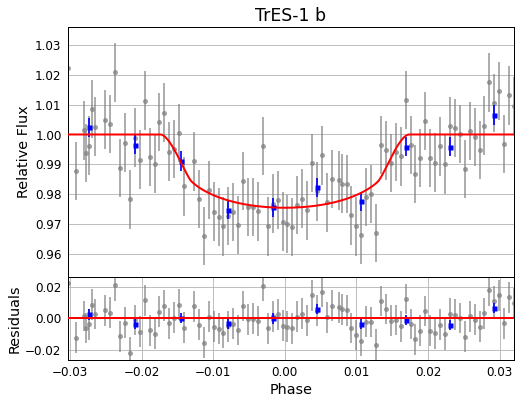}
}\subfloat[2018-07-19]{
\includegraphics[width=0.23\textwidth]{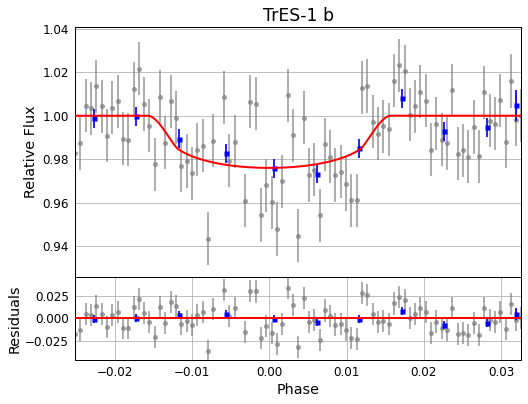}
}
\\
\subfloat[2018-08-04]{
\includegraphics[width=0.23\textwidth]{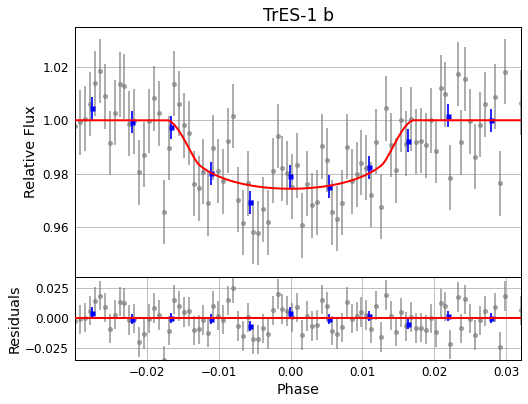}
}\subfloat[2018-10-18]{
\includegraphics[width=0.23\textwidth]{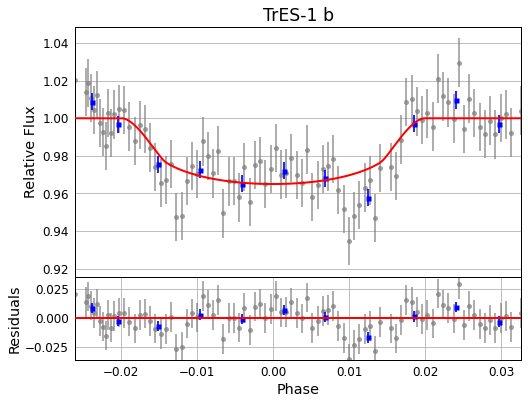}
}\subfloat[2018-10-21]{
\includegraphics[width=0.23\textwidth]{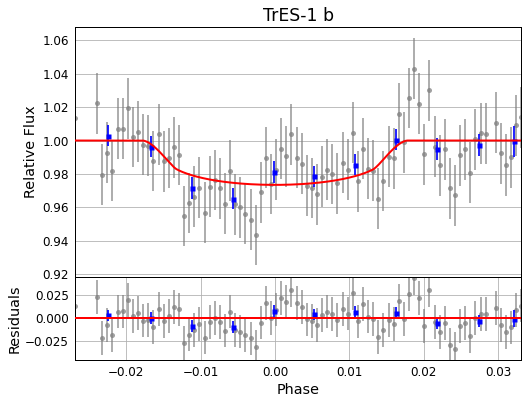}
}\subfloat[2019-05-21]{
\includegraphics[width=0.23\textwidth]{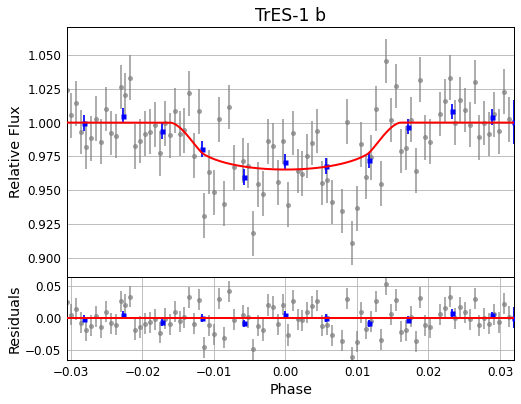}
}
\\
\subfloat[2019-05-27]{
\includegraphics[width=0.23\textwidth]{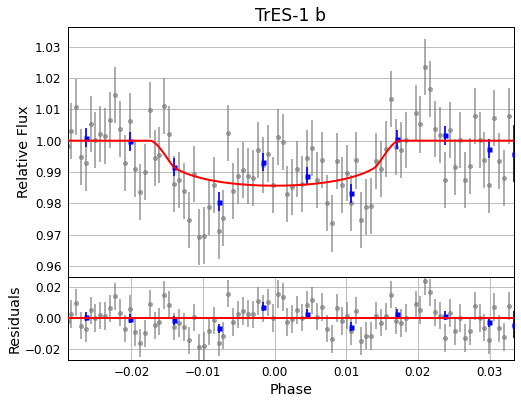}
}\subfloat[2019-06-03]{
\includegraphics[width=0.23\textwidth]{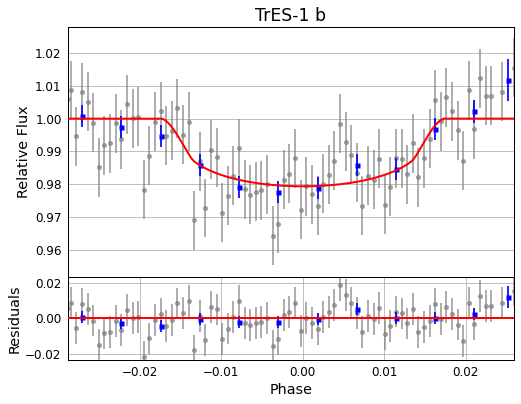}
}\subfloat[2019-08-20]{
\includegraphics[width=0.23\textwidth]{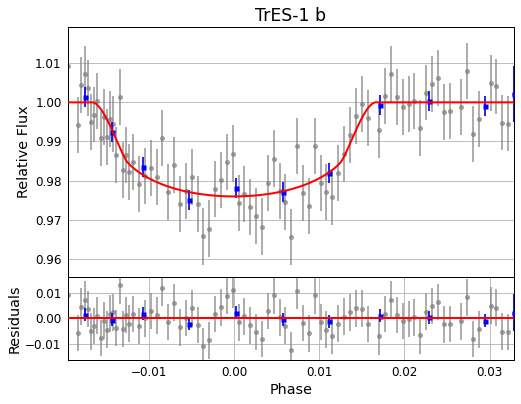}
}\subfloat[2019-08-26]{
\includegraphics[width=0.23\textwidth]{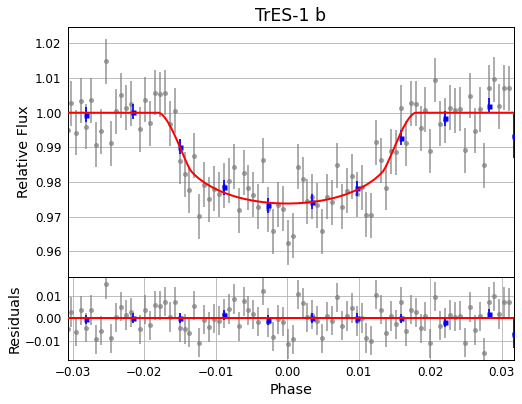}
}
\caption{Images of all good light curves. The date format is YYYY-MM-DD.}
\label{fig:transit pics}
\end{figure*}

\begin{figure*}[htbp]
\ContinuedFloat
\subfloat[2019-09-01]{
\includegraphics[width=0.23\textwidth]{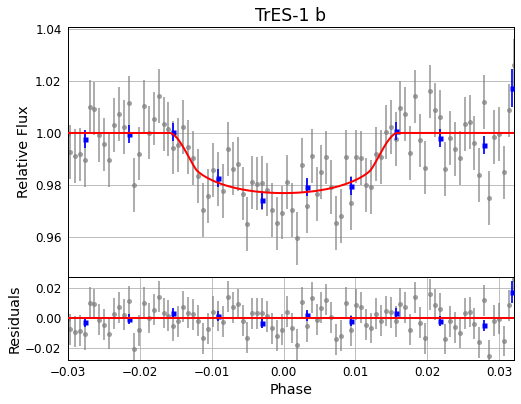}
}\subfloat[2019-09-04]{
\includegraphics[width=0.23\textwidth]{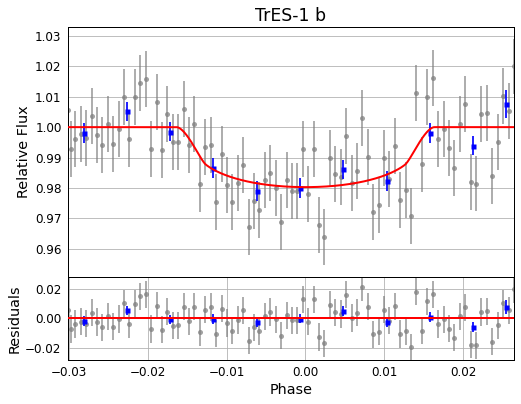}
}\subfloat[2020-06-28]{
\includegraphics[width=0.23\textwidth]{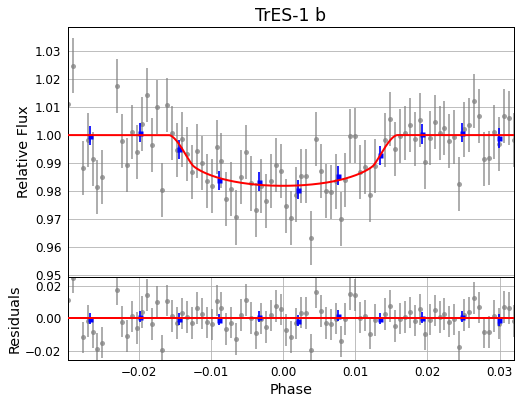}
}\subfloat[2020-07-04]{
\includegraphics[width=0.23\textwidth]{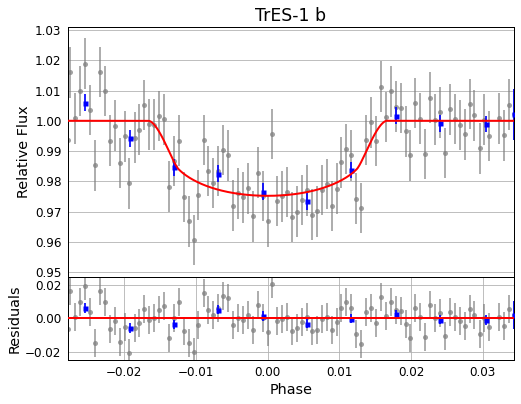}
}
\\
\subfloat[2020-09-27]{
\includegraphics[width=0.23\textwidth]{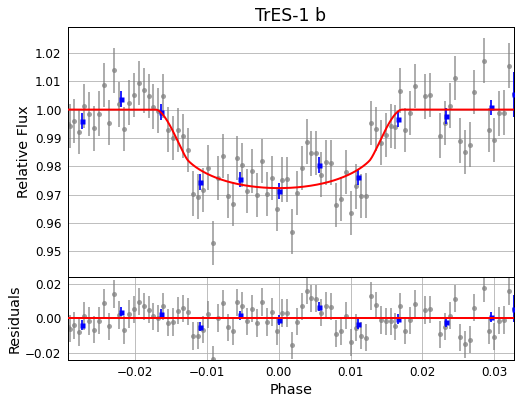}
}\subfloat[2020-09-30]{
\includegraphics[width=0.23\textwidth]{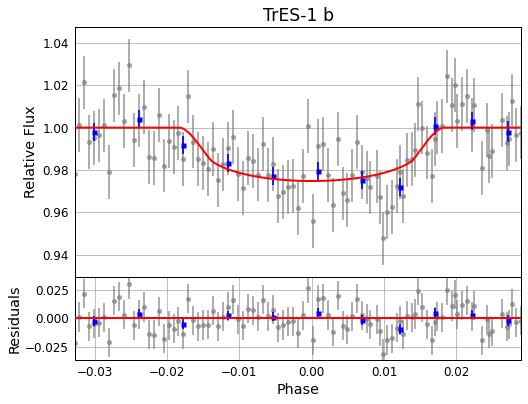}
}
\caption{Images of all good light curves. \textcolor{black}{The date format is YYYY-MM-DD.}}
\label{fig:transit pics}
\end{figure*}
\end{document}